\documentclass[prl,aps,twocolumn,showpacs]{revtex4}
\usepackage{epsfig,amssymb,amsfonts}
\newcommand\beq{\begin{equation}}
\newcommand\eeq{\end{equation}}
\newcommand\bea{\begin{eqnarray}}
\newcommand\eea{\end{eqnarray}}
\newcommand\non{\nonumber}
\newcommand\noi{\noindent}
\newcommand\al{\alpha}

\newcommand\om{\omega}

\newcommand\bib{\bibitem}
\newcommand\tdos{\textsf{TDOS~}}
\newcommand\tdosd{\textsf{TDOS}}
\newcommand\stm{\textsf{STM~}}
\newcommand\stmd{\textsf{STM}}

\newcommand\tll{\textsf{LL~}}
\newcommand\tlld{\textsf{LL}}
\newcommand\tlls{\textsf{LLs~}}

\newcommand\qwd{{\textsf{QW~}}}
\newcommand\qws{{\textsf{QWs}}}
\newcommand\fps{{\textsf{FPs}}}
\newcommand\fpsd{{\textsf{FPs~}}}
\newcommand\fp{{\textsf{FP}}}
\newcommand\fpd{{\textsf{FP~}}}
\newcommand\bc{{\textsf{BC}}}
\newcommand\bcd{{\textsf{BC~}}}

\newcommand\bcsd{{\textsf{BCs~}}}

\def\ket#1{| #1 \rangle}

\def\scxone#1{\langle #1 \rangle}

\def\me#1#2#3{\langle #2\, |~#1~|\, #3 \rangle}
\def\lsim{\mathrel{\rlap{\lower4pt\hbox{\hskip1pt$\sim$}}
\raise1pt\hbox{$<$}}} %less than or approx. symbol
\def\gsim{\mathrel{\rlap{\lower4pt\hbox{\hskip1pt$\sim$}}
\raise1pt\hbox{$>$}}} %greater than or approx. symbol
\newcommand{\re}[1]{\ensuremath{{\cal R}e #1}}

\def\dfrac#1#2{{\displaystyle\frac{#1}{#2}}}
\newcommand\dpfp{{${\mathsf{D_P}}$~}}

\newcommand\nfp{${\mathsf{N~}}$}

\newcommand\ie{{\textit{i.e.}}}

\begin{document}
\textheight=23.8cm

\title{\Large Enhancement of tunneling density of states at a junction of
three Luttinger liquid wires}
\author{\bf Amit Agarwal$^1$, Sourin Das$^1$, Sumathi Rao$^2$ and Diptiman
Sen$^1$}
\affiliation{\it $^1$ Centre for High Energy Physics, Indian Institute of
Science, Bangalore 560 012, India \\
\it $^2$ Harish-Chandra Research Institute, Chhatnag Road, Jhusi, Allahabad
211 019, India}
\date{\today}
%\pacs{71.10.Pm, 71.27.+a, 73.40.Gk, 61.16.Ch}

\begin{abstract}

We study the tunneling density of states (\tdosd) for a junction of three
Tomonaga-Luttinger liquid wires. We show that
there are fixed points which allow for the {\it enhancement} of the \tdosd,
which is unusual for Luttinger liquids. The distance from the junction over
which this enhancement occurs is of the order of $x = v/(2 \om)$, where $v$
is the plasmon velocity and $\om$ is the bias frequency. Beyond this distance,
the \tdos crosses over to the standard bulk value independent of the fixed
point describing the junction. This finite range of distances opens up the
possibility of experimentally probing the enhancement in each wire 
individually.
\end{abstract}

\maketitle

{\it Introduction.-} Junctions of multiple quantum wires (\qws) have attracted
considerable attention in recent years since they form the basic building 
blocks of quantum circuitry. Experimental realizations of three-wire 
$Y$-junctions of carbon nanotubes have also given the field a strong impetus 
\cite{fuhrerterrones}. Earlier studies of junctions of Tomonaga-Luttinger 
liquids (\tlld) were mostly focused on searching for various low-energy fixed 
points (\fps) 
\cite{nayak,lal,chen,affleck,meden,das2006drs,giuliano,bellazzini} and the
corresponding conductance matrices; this includes the recent development 
involving the inclusion of a finite superconducting gap at the junction 
\cite{drs2007,dasrao2008}.

Here we will focus on the local single-particle tunneling density of states
(\tdosd); this can be found by measuring the differential
tunneling conductance of a scanning tunneling microscope (\stmd) tip
at a finite bias. The differential conductance provides a direct
measure of the \tdos of the electrons at an energy given by the bias
voltage \cite{lltheory}, provided the density of states for the \stm is energy
independent. For a two-wire \tll junction, the current measured by the \stm
tip varies as a power of the bias \cite{eggert}; the power depends on the 
Luttinger parameter $g$ and the \fpd to which the junction has been tuned.

Earlier studies of the \tdos in a \tll system with \cite{WithImpurity} and 
without \cite{WithoImpurity} impurities revealed that the \tdos vanishes as 
a power law in the zero bias limit \footnote {An enhancement for the single
impurity case was observed in Ref. \cite{oreg} in a different context, which 
is outside the scope of this work.}. This is popularly considered to be a 
hallmark of a \tll system. The \tdos for a \tll wire with an impurity was 
studied in Ref. \cite {eggert} where the tunneling \stm current 
close to the impurity was shown to be power law suppressed in the zero bias 
limit. An enhancement of the \tdos was found at a single \tll-superconductor
\cite{hekking} junction and was explained in terms of the proximity effect.
An enhancement of the spectral weight was also found at a junction of multiple
\tlls tuned to a {\it fermionic} \fpd (linear boundary condition (\bc) between 
fermion fields at the junction) \cite{meden}. All the earlier studies of the 
\tdos were focused on either a multiple wire junction tuned to a fermionic \fpd
or a single \tll-superconductor junction, but not a junction of multiple \tlls
tuned to {\it bosonic} \fps. Bosonic \fpsd refer to linear \bcsd 
connecting incoming and outgoing currents (which are bilinears in the 
fermion fields) at the junction.

Here we study the \tdos of a three-wire junction of a
single channel \qwd modeled as a \tlld. We find that close to the
junction, the \tdos depends on the details of the current
splitting matrix at the junction which relates the incoming and 
outgoing currents; this matrix describes {\it bosonic} \fpsd
of the system. We find that for a certain range of
repulsive inter-electron interaction ($g<1$) and certain
current splitting matrices, the \tdos close to the junction shows an
{\it enhancement} in the zero bias limit. We show that this is
related to reflection of holes off the junction which mimics the
Andreev reflection process, even though there is no superconductor in
the present scenario. This is in sharp contrast to the case of a two-wire
junction where a repulsive electron-electron interaction always results
in a suppression of the \tdos near the junction. This is
the central result of this Letter. Far away from the junction, the \tdos
reduces to that of the bulk \tll wire, $\rho(\om) \sim \om^{(g+g^{-1}-2)/2}$,
independent of the details of the junction, and shows a suppression for
both repulsive ($g<1$) and attractive ($g>1$) interactions.

{\it Bosonization.-} The electron field (taken to be spinless for 
simplicity) can be bosonized as $\psi (x) =(1/\sqrt{2 \pi \alpha})$ $[F_O \,
e^{i k_F x + i \phi_{O}(x)} ~+~ F_I e^{-i k_F x + i \phi_{I}(x)} ]$,
\noi where $\phi_{O}(x)$, $F_O$ and $\phi_{I}(x)$, $F_I$ are the outgoing and
incoming chiral bosonic fields and corresponding Klein factors, $k_F$ is the 
Fermi momentum, and $\al$ is a short distance cut-off. We model the wires as 
spinless \tlls on a half-line ($x > 0$), \ie, all the wires are parametrized
by a coordinate $x$ running from $0$ to $\infty$. The corresponding 
Hamiltonian is given by 
\bea H &=& \int_0^\infty dx~ \sum_{i=1}^N~ \dfrac{v}{2\pi} ~\left\{ g \,
\left( \phi_i^\prime \right)^2 + \dfrac{1}{g}\, \left(\,\theta_i^\prime
\right)^2 \right\} , \label{hamiltonian} \eea
where prime (dot) stands for spatial (time)
derivative, $\phi_i (x,t) = (\phi_{Oi} + \phi_{Ii})/2$, $\theta_i
(x,t) = (\phi_{Oi} - \phi_{Ii})/2$, ${\dot \phi_i} =
(v/g)\,\theta_i^\prime$, and $ {\dot \theta_i} = (v
g)\,\phi_i^\prime$. The total density and current can be expressed
in terms of the incoming and outgoing fields as $\rho = \rho_O +
\rho_I$ with $\rho_{O/I} = \pm (1/2 \pi)\,\phi_{O/I}^\prime$, and $j
= j_O - j_I$ with $j_{O/I} = \pm (v_F/2\pi)\,\phi_{O/I}^\prime$.
Finally, we need to impose a boundary condition on the fields at $x=0$. 
Following Ref. \cite{das2006drs}, the incoming and outgoing currents, and 
consequently the bosonic fields, are related at the junction by a current 
splitting matrix $\mathbb{M}$, \ie, $j_{Oi} = \sum_j \mathbb{M}_{ij}~ 
j_{Ij}$, which leads to $\phi_{Oi} = \sum_j \mathbb{M}_{ij} ~\phi_{Ij}$.
% , where we have ignored an integration constant.
To ensure that the matrix $\mathbb{M}$ represents a \fpd of the theory, the 
incoming and outgoing fields must satisfy the bosonic commutation 
relations; this restricts the matrix $\mathbb{M}$ to be orthogonal. Scale 
invariance imposes the same constraint of orthogonality on $\mathbb{M}$ as 
shown in Ref. \cite{dasrao2008}; orthogonality also implies that there is no 
dissipation in the system. Current conservation at the junction implies 
\cite{bellazzini} that each row and column of $\mathbb{M}$ adds up to 1.

In general, for a three-wire charge-conserving junction, 
$\mathbb{M}$ can be parametrized by a single continuous parameter $\theta$,
and it falls into one of two classes with (a) $ \det \mathbb{M}_1 = 1$, and 
(b) $\det \mathbb{M}_2=-1$. These classes can be expressed as
\beq \label{m1} \mathbb{M}_1 = \left(\begin{array}{ccc}
a & b & c \\
c & a & b \\
b & c & a \end{array}\right), \quad \mathbb{M}_2 = \left(\begin{array}{ccc}
b & a & c \\
a & c & b \\
c & b & a \end{array}\right). \eeq
In Eq. (\ref{m1}), $a=(1+2\cos\theta)/3$, $b(c)=(1-\cos \theta
+(-) \sqrt{3} \sin \theta)/3$.
This provides us with an explicit parametrization of the two
one-parameter families of \fps; any \fpd in the
theory can be identified in terms of $\theta$, with the \fpsd
at $\theta=0$ and $\theta=2 \pi$ being identical. The $\det
\mathbb{M}_1 = 1$ class represents a $Z_3$ symmetric (in the wire
index) class of \fps, while $\det \mathbb{M}_2 = -1$
represents an asymmetric class of \fps. In the
$\mathbb{M}_1$ class, $\theta = \pi$ corresponds to the \dpfp \fp,
$\theta=0$ corresponds to the disconnected \nfp \fp, and $\theta = 2\pi/3$ 
and $4\pi/3$ correspond to the chiral cases $\chi_{\pm}$, following the 
notation of Ref. \cite{affleck}. Since $\phi_{O}$ 
and $\phi_{I}$ are interacting fields, we must perform a Bogoliubov
transformation, $\phi_{O/I}=(1/2{\sqrt g})\{(1 +
g)\tilde{\phi}_{O/I} + (1 - g){\tilde {\phi}_{I/O}} \}$, to obtain
the corresponding free outgoing (incoming) ($\tilde\phi_{O/I}$)
chiral fields satisfying the commutation relations,
$[\tilde{\phi}_{O/I}(x,t),\tilde{\phi}_{O/I}(x',t)]=\pm i\pi
Sgn(x-x')$. Unlike the usual Bogoliubov transformation,
here we also need to consider the effect of the junction matrix 
$\mathbb{M}$ relating the interacting incoming and outgoing
fields. Following Ref. \cite{das2006drs}, we obtain a
Bogoliubov transformed matrix $\widetilde{\mathbb{M}}$ which relates
$\tilde{\phi}_{Oi}$ to $\tilde{\phi}_{Ii}$. We find that
$\tilde{\phi}_{Oi} (x) = \sum_j ~ \widetilde{\mathbb{M}}_{ij}
~\tilde{\phi}_{Ij} (-x)$ where $\widetilde{\mathbb{M}} = $ $
\left[(1+g)\mathbb{I}-(1-g)\mathbb{M}\right]^{-1}
\left[(1+g)\mathbb{M}-(1-g)\mathbb{I}\right]$. Note that the $\mathbb{M}_2$
class of matrices satisfy $(\mathbb{M}_2)^2=\mathbb{I}$; hence
$\widetilde{\mathbb{M}}_2=\mathbb{M}_2$, and the interacting and free 
fields satisfy identical \bcsd at the junction. This is not true for the 
$\mathbb{M}_1$ class, but $\widetilde{\mathbb{M}}_1$ still has the same form 
as $\mathbb{M}_1$ with the corresponding parameters $\tilde a = {(3g^2-1 + 
(3g^2+1)\cos{\theta})}/{\eta}$ and
$\tilde b (\tilde c) = $ ${2(1 - \cos{\theta} +(-) \sqrt{3} g \sin{\theta} )}/
\eta$, where $\eta={3(1+g^2+ (g^2-1)\cos{\theta})}$. 
% Note that the matrices $\widetilde{\mathbb{M}}_1$ 
% are non-linear functions of the \tll parameter $g$,
% while the matrices $\widetilde{\mathbb{M}}_2$ are independent of $g$.

\begin{figure}[t]
\begin{center} \epsfig{figure=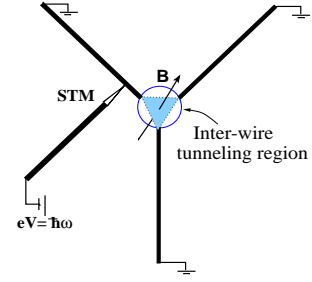,width=4cm,height=3.7cm} \end{center}
\vskip -.5cm
\caption{Schematic picture of STM tip for measuring the TDOS near the 
junction, the region of inter-wire tunnelings and a magnetic field $\bf B$
at the junction.} \label{spfig2} 
\vskip -.4cm
\end{figure}

The $\mathbb{M}$ matrix 
% is not just a mathematical construct but 
is related to the DC conductance matrix given by ${\mathbb{G}}
\label{GL}~=~ (2e^2/h) ({\mathbb{I}}~-~\mathbb{M})$ for Fermi liquid leads 
\cite{affleck,dasrao2008}. Qualitatively, $\mathbb{M}$ is related to 
tunnelings between the different wires and backscatterings in each wire. The 
experimental set-up can be a junction of several edges of a quantum Hall 
system as in Ref. \cite{das2006drs}, and $\mathbb{M}$ (or $\theta$) can 
be tuned by applying gate voltages and a magnetic field at the junction. We 
note that $\mathbb{M}_2$ is time-reversal invariant, but $\mathbb{M}_1$ is
generally not and tuning it will require a magnetic field piercing the 
junction (see Fig. \ref{spfig2}). 

{\it Tunneling density of states.-} We now compute the \tdos for adding an 
electron with energy $\hbar \om$ on the $i^{\rm th}$ wire \cite{lltheory},
\vskip -0.5cm \bea \rho_i (\om) &=& 2 \pi \sum_n \vert
\me{\psi_i^\dagger (x)}{0}{n} \vert^2 ~\delta (E_n - E_0 - \hbar \om) \non \\
&=& 2 \re{\int_0^\infty dt ~\me{\psi_i (x,t)\, \psi_i^\dagger(x,0)}{0}{0} ~
e^{i\om t}}. \label{tdosdef} \eea \vskip
-0.2cm \noi Here $\ket{n}$ ($E_n$) denotes the $n^{\rm th}$ eigenstate 
(eigenvalue) of the Hamiltonian in Eq. (\ref{hamiltonian}).
% with an appropriate boundary condition. 
% and $E_n$ is the corresponding eigenvalue. 
The Green's function in the $i^{\rm th}$ wire is
${\cal G} = \scxone{\psi_i (x,t) \psi^{\dagger}_i(x,0)} = \scxone{
\psi_{Ii}(x,t) \psi^{\dagger}_{Ii}(x,0)} + \scxone{\psi_{Oi} (x,t)
\psi^{\dagger}_{Oi}(x,0)} + e^{-2ik_F x} \scxone{ \psi_{Ii} (x,t)
\psi^{\dagger}_{Oi}(x,0)} + e^{2ik_F x}$ $\scxone{\psi_{Oi} (x,t)
\psi^{\dagger}_{Ii}(x,0)}$. The two non-oscillatory terms are 
\bea \label{nofriedel}
& & \scxone{\psi_{Ii}^{} (x,t) \,\psi^{\dagger}_{Ii}(x,0)} =
\scxone{\psi_{Oi}^{} (x,t) \,\psi^{\dagger}_{Oi}(x,0)} \non \\
&=& \dfrac{1}{2\pi \al} \left[\dfrac{i\al}{-vt+i\al}
\right]^{\frac{(1+g^2)}{2g}} \left[\frac{-\al^2-4x^2}{(-vt+i\al)^2
- 4x^2} \right]^{\frac{\tilde d_i (1-g^2)}{4g}}. \eea
The oscillatory part vanishes as $L \to \infty$ and can 
be dropped in further discussions. For the $\widetilde{\mathbb{M}}_1$ class, 
$\tilde d_i = \tilde a$; for the $\widetilde{\mathbb{M}}_2$ class,
$\tilde d_i={\tilde a}, {\tilde b},{\tilde c}$ depending on the wire index.

\begin{figure}[t]
\begin{center} \epsfig{figure=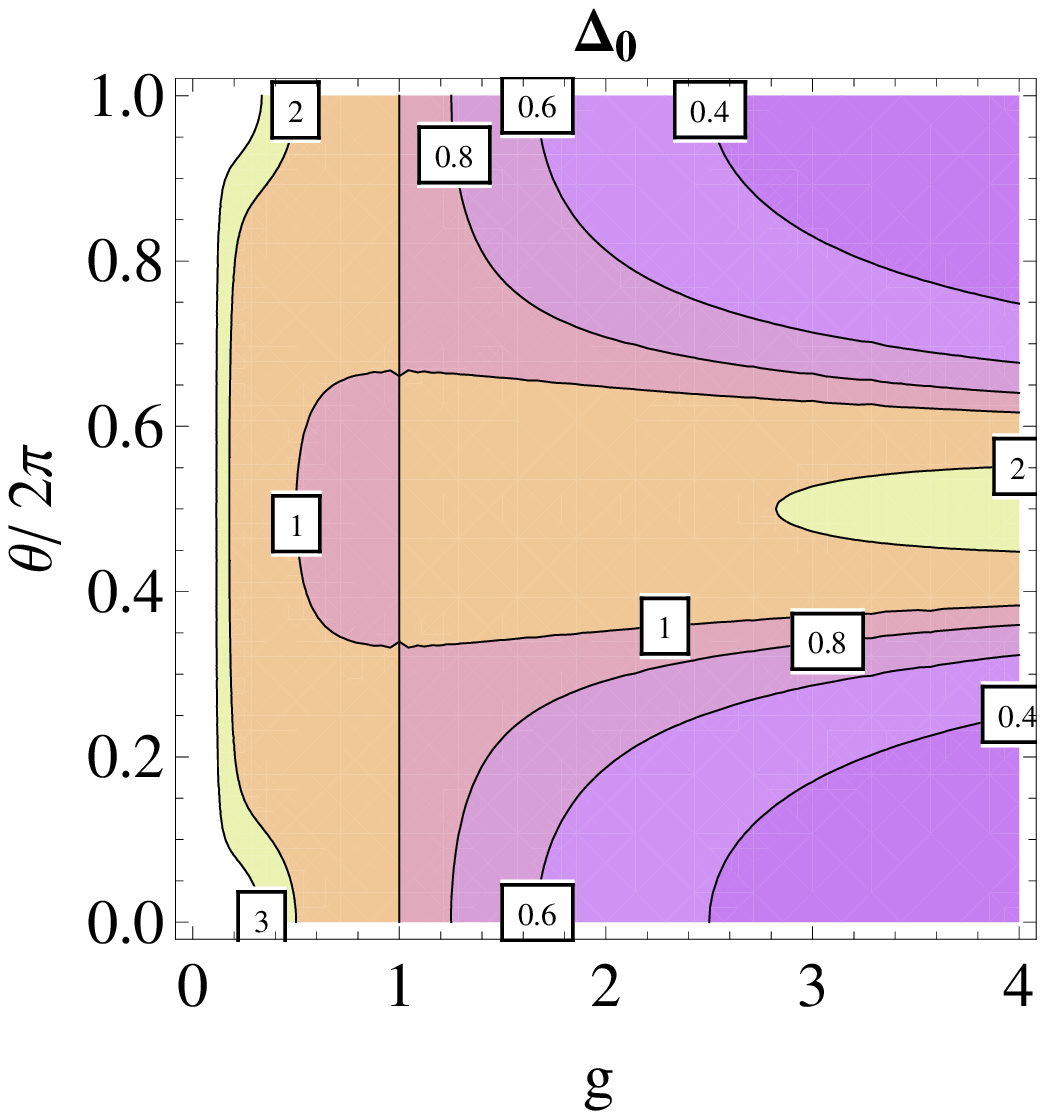,width=4cm,height=4cm}
\epsfig{figure=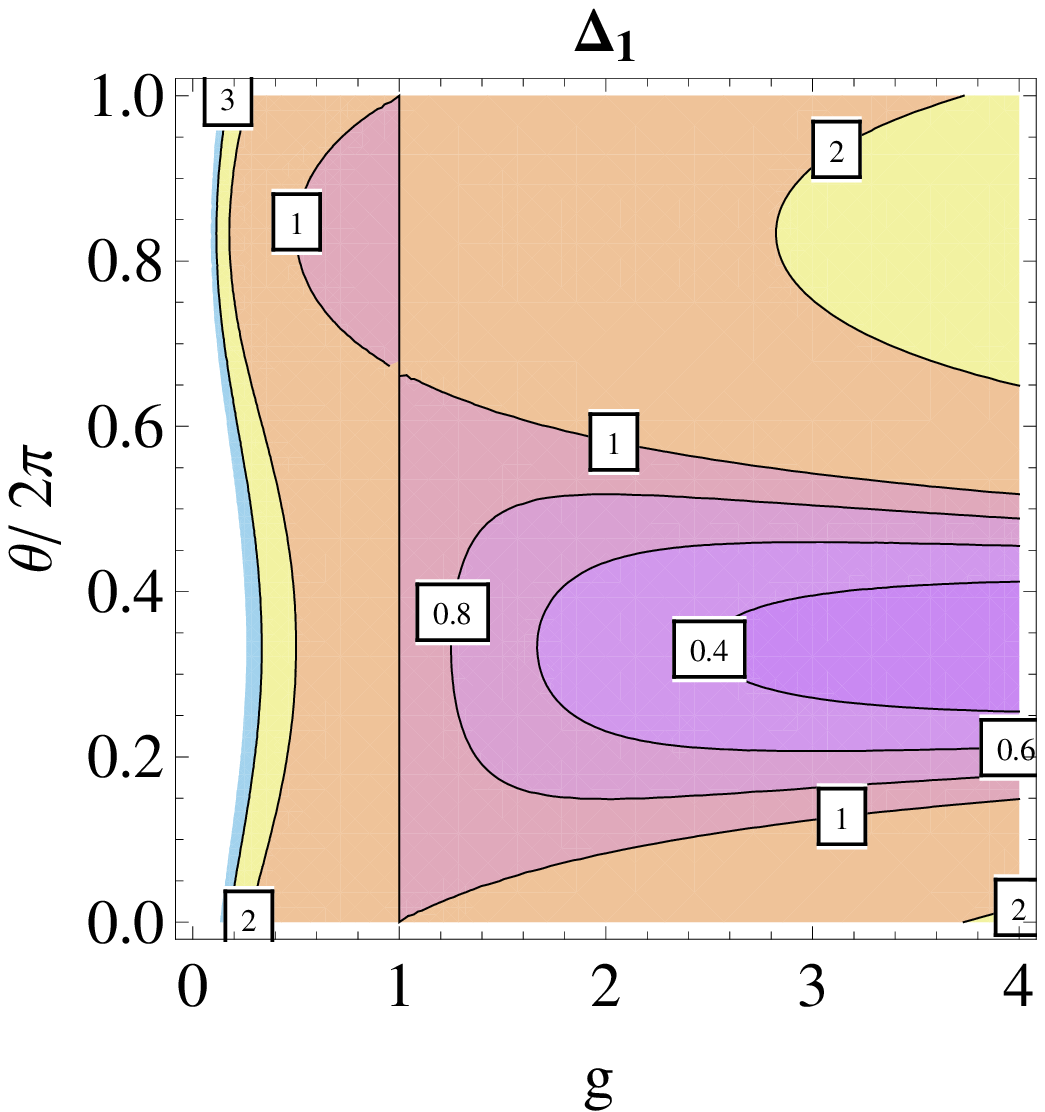,width=4cm,height=4cm} \end{center}
\vskip -.5cm
\caption{$\Delta_0$ for all wires in the $\mathbb{M}_1$ class (left), and
$\Delta_1$ for wire 1 in the $\mathbb{M}_2$ class (right) in the $g-\theta$ 
plane.} 
%$\Delta_2$, $\Delta_3$ for wires 2, 3 in the $\mathbb{M}_2$ class are 
%obtained by shifting $\theta$ to $\theta - 2\pi/3$ and $\theta + 2\pi/3$ 
%respectively in the right picture.} 
\label{spfig1} 
\vskip -.4cm
\end{figure}

Treating the tunneling strength $\gamma$ between the $i^{\rm {th}}$ wire and 
the \stm tip perturbatively and using Eqs.
(\ref{tdosdef}-\ref{nofriedel}), the differential tunneling conductance
evaluated to leading order in $\gamma$ is found to be directly proportional
\cite{lltheory} to the \tdos on the $i^{\rm {th}}$ wire. The \tdos has the 
same form in the $x \to 0 $ and $x \to \infty $ limits and is given by
\beq \rho_i (\om) = \dfrac{1}{\alpha \hbar{{\bf{\Gamma}}(\Delta)}}~
{{\tau}_c}^{\Delta}~{\om}^{\Delta-1} ~e^{{-|\om| \al/v}} ~\Theta(\om),
\label{tdos} \eeq
where ${\bf{\Gamma}}(\Delta)$ is the Gamma function, $\Theta(\om)$ is the 
Heaviside step function, $\tau_c =\al/v$ is the short time cut-off, and $\om 
= e V /\hbar$, where $e$ is the electronic charge and $V$ is the bias voltage
between the \stm tip and wire system held at a uniform potential. The cut-off
frequency scale for the validity of the perturbation theory is given
by $\om_0=[{|\gamma|^{-2/(\Delta_i-1)}v}]/{\al}$. For $x \to 0$,
$\Delta$ is a function of the $\tilde d_i$ which is the corresponding
diagonal element of the appropriate $\widetilde{\mathbb{M}}$ matrix.
For the $\mathbb{M}_1$ class, $\Delta = \Delta_0 (\tilde a)$, while
for $\mathbb{M}_2$, $\Delta = \Delta_i$ is a function of
${\tilde a},{\tilde b},{\tilde c}$ depending on the wire index $i$.
For $x \to \infty$, $\Delta =(g+g^{-1})/2$ independent of the
$\mathbb{M}$ matrix at the junction. Thus we recover the expression for the 
bulk \tdos in a $\tll$ as $x \to \infty$ irrespective of the details of the 
junction. For the $\mathbb{M}_1$ class, the power law exponent for $x \to 0$ 
is the same on all the three wires due to $Z_3$ symmetry and is given by
\beq \Delta_0 = \frac{1}{3g}~\frac{5g^2+1+(g^2-1)~\cos{\theta}}{g^2+1+
(g^2-1)~\cos{\theta}}. \label{deli} \eeq
Eq. (\ref{deli}) indicates that for $g<1$ there are values of
$\theta$ for which $\Delta < 1$. This implies that there are \fpsd
in the theory which show an {\it enhancement} (see Eq. (\ref{tdos}))
of the \tdos in the zero bias limit $\om \to 0$. This is in sharp
contrast to previous studies for various cases of normal (not
superconducting) junctions of two-wire systems which always showed a
suppression of the \tdos in the zero bias limit for $g<1$. This is
our main result. Note that whenever $g=1$, $\Delta=1$; this
implies that the \tdos is independent of the bias for $g=1$.
This might look natural since $g=1$ corresponds to free fermions in
the wire for which the \tdos is expected to be bias (energy) independent.
However, this is misleading; the \bcd conditions expressed in terms of the 
matrix $\mathbb{M}$ at the junction correspond to non-linear
relations between the fermions on each wire in the vicinity of the
junction, and hence represent non-trivial interaction between the
fermions at the junction. Hence, the \tdos being energy independent
for $g=1$ for any \fpd represented by $\mathbb{M}_1$ is a
non-trivial result by itself. To get a clear idea about the \fpsd
which show an enhancement of the \tdos, we present contour plots of
$\Delta$ in the $g-\theta$ plane in Fig. \ref{spfig1}. In the left plot 
(corresponding to the $\mathbb{M}_1$ class), we can see a dome shaped region 
for $g < 1$ which corresponds to \fpsd showing an enhancement, \ie, 
$\Delta_0 <1$. It is interesting to note that this region is bounded by the 
two chiral \fps, $\chi_{\pm}$ at $\theta = 2\pi/3, ~4\pi/3$. The 
\dpfp \fpd at $\theta=\pi$ also falls in this region and shows an 
enhancement of the \tdos for $1/2 < g < 1$. For the $\mathbb{M}_2$ class, 
the power law exponents for the three wires are given by
\bea \Delta_1 &=& \frac{4+2g^2+(\cos{\theta}-\sqrt{3}\sin{\theta})~
(g^2-1)}{6g}, \eea
$\Delta_2$ and $\Delta_3$ which are obtained by shifting $\theta \to \theta 
\mp 2\pi/3$ in $\Delta_1$. For this class also, there are \fpsd which 
show an enhancement of the \tdos for $g<1$; they correspond to the dome 
shaped region in the right plot in Fig. \ref{spfig1}. In contrast to the 
$\mathbb{M}_1$ class, in this case there can be an enhancement in one wire 
and suppression in the other wires due to the broken $Z_3$ symmetry. 

\begin{figure}[t]
\begin{center} \epsfig{figure=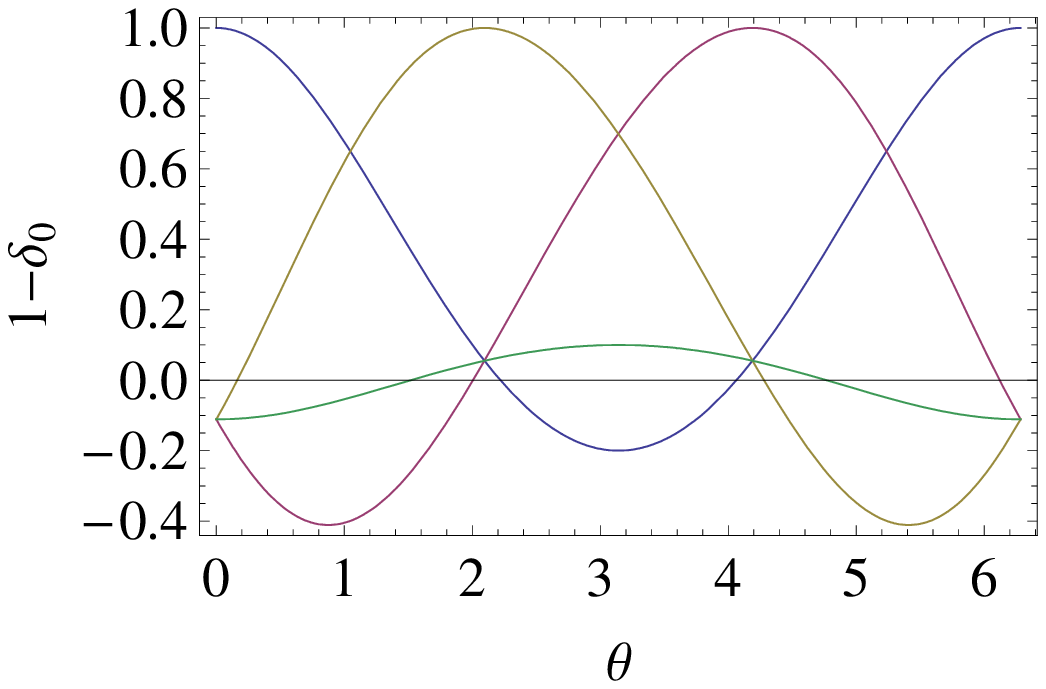,width=4cm,height=4cm}
\epsfig{figure=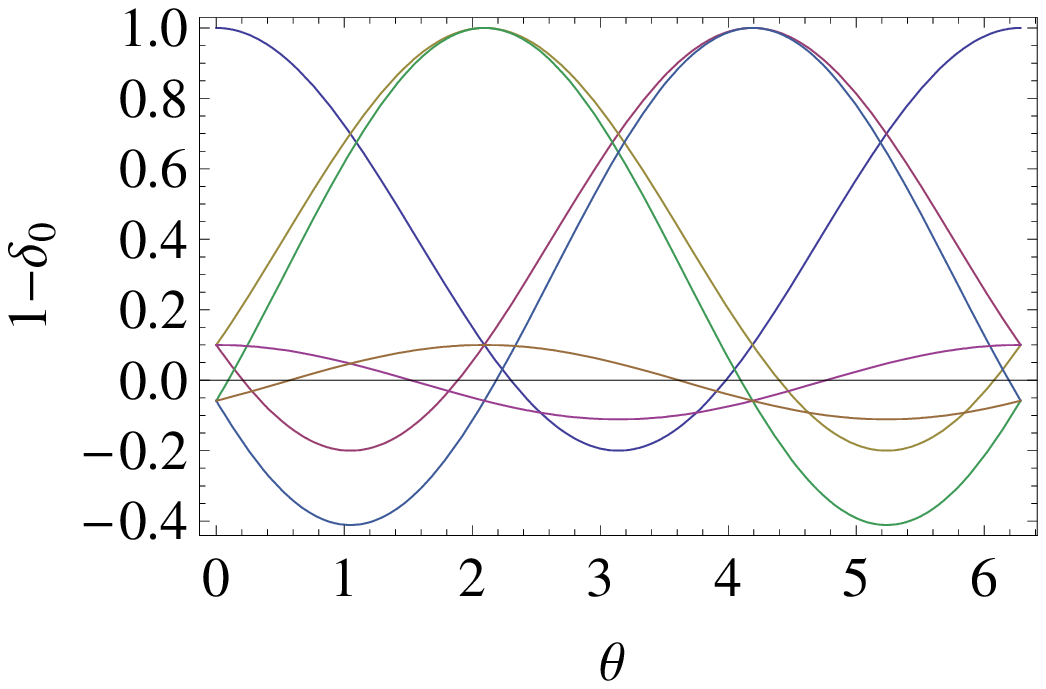,width=4cm,height=4cm} \end{center}
\vskip -.5cm
\caption{$1-\delta_0$ as functions of $\theta$ for the classes $\mathbb{M}_1$
(left) and $\mathbb{M}_2$ (right) for various tunneling operators, for 
$g=0.9$.} \label{spfig3} 
\vskip -.4cm
\end{figure}

The observability of the different \fpsd and the enhancement of the 
TDOS near the junction crucially depends on the renormalization group 
(RG) stability of the \fpd against various perturbations in the form of 
inter-wire electron tunnelings at the junction (see Fig. \ref{spfig2}). 
To understand the stability of the \fp, we present the scaling dimension 
$\delta_0$ for all possible single-electron tunneling events (quadratic 
in fermion operators) in Tables \ref{operator1} and \ref{operator2}; in these 
tables, $\psi_{I/O}$ are related to $\phi_{I/O}$ by the bosonization formula 
given earlier. In Fig. \ref{spfig3}, we plot the values of $1-\delta_0$ as 
functions of $\theta$ for $g=0.9$. A positive value of $1-\delta_0$ implies 
that the operator is RG relevant. 

\begin{table}[htb]
\begin{center}
\begin{tabular}{c c}
\hline \hspace{0.2cm} {\sf {Operator}} \hspace{0.2cm} &
\hspace{0.2cm} {\sf {Scaling dimension $\delta_0$}} \hspace{0.2cm} \\ \hline
\hspace{0.5cm} $\psi_{iO}^{\dagger} \psi_{iI}^{}$ & $\frac{4g(1-\cos
\theta)}{3(g^2+\cos\theta(g^2-1)+1)}$ \\
\hspace{0.5cm} $\psi_{2O}^{\dagger} \psi_{1I}^{} , \psi_{3O}^{\dagger} 
\psi_{2I}^{}, \psi_{1O}^{\dagger} \psi_{3I}^{}$ & $\frac{2g(\cos\theta+
\sqrt{3}\sin\theta+2)}{3(g^2+\cos\theta(g^2-1)+1)}$ \\
\hspace{0.5cm}$ \psi_{1O}^{\dagger} \psi_{2I}^{} , \psi_{2O}^{\dagger} 
\psi_{3I}^{} , \psi_{3O}^{\dagger} \psi_{1I}^{}$
& $\frac{2g(\cos\theta-\sqrt{3}\sin\theta+2)}{3(g^2+\cos\theta(g^2-1)+1)}$ \\
\hline
\hspace{0.5cm} $\psi_{2I}^{\dagger} \psi_{1I}^{} , \psi_{3I}^{\dagger} 
\psi_{2I}^{}, \psi_{1I}^{\dagger} \psi_{3I}^{} $ & $\frac{2g}{g^2+\cos
\theta(g^2-1)+1} $ \\
\hspace{0.5cm}$ \psi_{2O}^{\dagger} \psi_{1O}^{} , \psi_{3O}^{\dagger} 
\psi_{2O}^{}, \psi_{1O}^{\dagger} \psi_{3O}^{}$ & $\frac{2g}{g^2+\cos\theta
(g^2-1)+1}$ \\
\hline
\end{tabular}
\caption[]{\footnotesize{Table of tunneling operators for the $\mathbb{M}_1$}
class.}
\label{operator1}
\end{center}
\end{table}
\vskip -.5cm

\begin{table}[htb]
\begin{center}
\begin{tabular}{c c}
\hline \hspace{0.2cm} {\sf {Operator}} \hspace{0.2cm} &
\hspace{0.2cm} {\sf {Scaling dimension $\delta_0$}} \hspace{0.2cm} \\ \hline
\hspace{0.5cm} $\psi_{1O}^{\dagger} \psi_{1I}^{}$ & $\frac{1}{3} g(2-2\cos
\theta)$ \\
\hspace{0.5cm} $\psi_{2O}^{\dagger} \psi_{2I}^{}$ & $\frac{1}{3} g(2+\cos
\theta + {\sqrt 3} \sin \theta)$ \\
\hspace{0.5cm} $\psi_{3O}^{\dagger} \psi_{3I}^{}$ & $\frac{1}{3} g(2+\cos
\theta - {\sqrt 3} \sin \theta)$ \\
\hline
\hspace{0.5cm}$ \psi_{1O}^{\dagger} \psi_{2I}^{} , \psi_{2O}^{\dagger} 
\psi_{1I}^{}$ & $\frac{3+g^2}{12 g} (2-2\cos\theta )$ \\
\hspace{0.5cm}$ \psi_{2O}^{\dagger} \psi_{3I}^{} , \psi_{3O}^{\dagger} 
\psi_{2I}^{}$ & $\frac{3+g^2}{12 g} (2+\cos\theta -{\sqrt 3} \sin\theta)$ \\
\hspace{0.5cm}$ \psi_{3O}^{\dagger} \psi_{1I}^{} , \psi_{1O}^{\dagger} 
\psi_{3I}^{} $ & $\frac{3+g^2}{12 g} (2+\cos\theta +{\sqrt 3} \sin\theta)$ \\
\hline
\hspace{0.5cm} $\psi_{1I}^{\dagger} \psi_{2I}^{} , \psi_{1O}^{\dagger} 
\psi_{2O}^{}$ & $\frac{2(g^2+1)+(g^2-1)2 \cos\theta}{4g}$ \\
\hspace{0.5cm} $\psi_{2I}^{\dagger} \psi_{3I}^{} , \psi_{2O}^{\dagger} 
\psi_{3I}^{}$ & $\frac{2(g^2+1)-(g^2-1)(\cos\theta -\sqrt{3} \sin\theta)}{4g}$
\\
\hspace{0.5cm} $\psi_{3I}^{\dagger} \psi_{1I}^{} , \psi_{3O}^{\dagger} 
\psi_{1I}^{}$ & $\frac{2(g^2+1)-(g^2-1)(\cos\theta +\sqrt{3} \sin
\theta)}{4g}$ \\ 
\hline
\end{tabular}
\caption[]{\footnotesize{Table of tunneling operators for the ${\mathbb{M}_2}$}
class.} \label{operator2}
\end{center}
\end{table}
\vskip -.1cm

It is clear from Fig. \ref{spfig3} that all the \fpsd showing 
an enhancement of the \tdos are unstable; 
% for both the $\mathbb{M}_1$ and $\mathbb{M}_2$ classes; 
however, for the $\mathbb{M}_1$ 
class close to $\theta= 2\pi/ 3, ~4\pi/3$ (the $\chi_{\pm}$ \fps), only one 
operator is highly relevant and the rest are almost marginal. Hence this part 
of the parameter space allows for a large temperature window for observing
an enhancement if one can experimentally suppress the most 
relevant tunneling by tuning the junction appropriately.
% to the vicinity of this \fp.

To gain a better understanding of the enhancement, we expand the $\Delta$'s
for $g \simeq 1$ in the small parameter $1-g$ to obtain $\Delta_j = 1 + (1-g)
d_j$, where $j=0,1,2,3$, and $d_j$ are the diagonal elements of the
corresponding $\mathbb{M}$ matrix. This limit corresponds to weakly
interacting electrons in the bulk of the wires away from the junction.
Whenever $d_j<0$ and $g<1$, $\Delta_j$ is less then unity
which corresponds to an enhancement of the \tdos in the zero bias
limit. But $d_j < 0$ corresponds to a hole current being
reflected from the junction when an electron current is
incident on the junction. Hence we conclude that all the \fpsd 
which involve reflection of a hole off the junction lead to
an enhancement of the \tdosd. As discussed earlier, an enhancement of
the \tdos was previously observed in a junction of a
\tll wire with a superconductor \cite{hekking}; this can be attributed
to the proximity induced Andreev process at the junction
which results in the reflection of a hole from the junction in
response to an incident electron. It is interesting to note that for
our case too, the enhancement is connected to holes being reflected
off the junction, even though there is no superconductor in our
model. Finally, note from Eqs. (\ref{nofriedel}-\ref{tdos}) that
the cross-over length scale beyond which the \tdos goes over to its bulk 
form is given by $x = v/(2\om)$. For a typical bias voltage of $10 ~\mu V$ 
and a Fermi velocity $v\approx 10^5$ m/s (typical of a two-dimensional
electron gas), we get a cross-over length of about $3 ~\mu m$; this is 
readily accessible within present day experimental realizations of a 
one-dimensional \qwd in a two-dimensional electron gas.

{\it Discussion.-} It is important to note that the interesting prediction of
an enhancement of the \tdos for $g<1$ involves unstable \fps; hence
the enhancement can be expected to be observed in experiments at high 
temperatures only. If the junction is tuned to one of the \fpsd 
which show enhancement, a variation of the temperature from high to low 
will first show an enhancement and then a suppression of the \tdos as the 
system finally flows to the disconnected stable \fpd at low temperatures. 
This non-monotonicity observed via the \stm current will be a hallmark 
of our prediction. Here, high and low temperatures are defined with respect 
to a cross-over scale called $\omega_0$ after Eq. (\ref{tdos}). In other 
Luttinger liquid systems, such as the one studied experimentally in Ref. 
\cite{roddaro}, the cross-over scale $T_B$ was found to be of the order
of $0.5 - 3 K$ (corresponding to $\omega_0 \sim k_B T_B /\hbar \sim 60 - 400 
GHz$); this scale can be varied by tuning the tunneling strength $\gamma$. 
Thus the cross-over scale can be tuned experimentally as was done in Ref. 
\cite{roddaro}. Hence the temperature window in which the enhancement of 
the \tdos can be observed is experimentally tunable.

\end{document}